\documentclass[conference,letterpaper]{IEEEtran}
\IEEEoverridecommandlockouts
\usepackage[hidelinks, bookmarks=false]{hyperref}
\usepackage{cite}
\usepackage{amsmath,amssymb,amsfonts}
\usepackage{graphicx}
\usepackage{textcomp}
\usepackage{xcolor}
\usepackage{hyperref}
\usepackage{multirow}
\usepackage{threeparttable}
\usepackage{booktabs}
\usepackage{colortbl}
\definecolor{lightgray}{gray}{0.9}
\usepackage{algorithm}
\usepackage{algpseudocode}
\def\BibTeX{{\rm B\kern-.05em{\sc i\kern-.025em b}\kern-.08em
    T\kern-.1667em\lower.7ex\hbox{E}\kern-.125emX}}

\makeatletter
\def\ps@headings{%
\def\@oddhead{\mbox{}\scriptsize\rightmark \hfil \thepage}%
\def\@evenhead{\scriptsize\thepage \hfil \leftmark\mbox{}}%
\def\@oddfoot{}%
\def\@evenfoot{}}
\makeatother
\begin{document}

\title{Safeguarding Federated Learning-based \\ Road Condition Classification
}

\author{\IEEEauthorblockN{Sheng Liu}
\IEEEauthorblockA{\textit{Networked Systems Security Group} \\
\textit{KTH Royal Institute of Technology}\\
Stockholm, Sweden \\
shengliu@kth.se}
\and
\IEEEauthorblockN{Panos Papadimitratos}
\IEEEauthorblockA{\textit{Networked Systems Security Group} \\
\textit{KTH Royal Institute of Technology}\\
Stockholm, Sweden \\
papadim@kth.se}
}

\maketitle

\begin{abstract}
Federated Learning (FL) has emerged as a promising solution for privacy-preserving autonomous driving, specifically camera-based Road Condition Classification (RCC) systems, harnessing distributed sensing, computing, and communication resources on board vehicles without sharing sensitive image data. However, the collaborative nature of FL-RCC frameworks introduces new vulnerabilities: Targeted Label Flipping Attacks (TLFAs), in which malicious clients (vehicles) deliberately alter their training data labels to compromise the learned model inference performance. Such attacks can, e.g., cause a vehicle to mis-classify slippery, dangerous road conditions as pristine and exceed recommended speed. However, TLFAs for FL-based RCC systems are largely missing. We address this challenge with a threefold contribution: 1) we disclose the vulnerability of existing FL-RCC systems to TLFAs; 2) we introduce a novel label-distance-based metric to precisely quantify the safety risks posed by TLFAs; and 3) we propose FLARE, a defensive mechanism leveraging neuron-wise analysis of the output layer to mitigate TLFA effects. Extensive experiments across three RCC tasks, four evaluation metrics, six baselines, and three deep learning models demonstrate both the severity of TLFAs on FL-RCC systems and the effectiveness of FLARE in mitigating the attack impact.
\end{abstract}


\section{Introduction}
\textbf{Road Condition Classification (RCC)} \cite{zhao2023comprehensive}, encompassing tasks such as unevenness detection, friction estimation, and surface material identification, is important for intelligent transportation. It directly influences vehicle control, traffic safety, and passenger comfort. Based on the significant and growing deployment of on-board vehicle cameras and thanks to the rapid development of Artificial Intelligence (AI), Deep Neural Network (DNN) models can be efficiently trained and then deployed on vehicles to support RCC tasks based on images captured by onboard cameras \cite{10705359}. 


High-performance RCC models demand extensive training data covering diverse scenarios (e.g., varying weather, road types, lighting conditions) to ensure robustness \cite{YUAN2021385}. Nonetheless, escalating concerns on user privacy, e.g., GDPR\footnote{https://gdpr-info.eu/} in Europe, CCPA\footnote{https://www.oag.ca.gov/privacy/ccpa} in United States, and PIPL\footnote{http://en.npc.gov.cn.cdurl.cn/2021-12/29/c\_694559.htm} in China, render the traditional \textbf{centralized learning paradigm} impractical for training and updating RCC models in the long term. Moreover, adversaries can easily exploit vehicle or even passenger private details: e.g., the recently proposed GeoSpy\footnote{https://geospy.ai/} can predict the location of photos within seconds.

\begin{figure}[t]
\centerline{\includegraphics[width=0.45\textwidth]{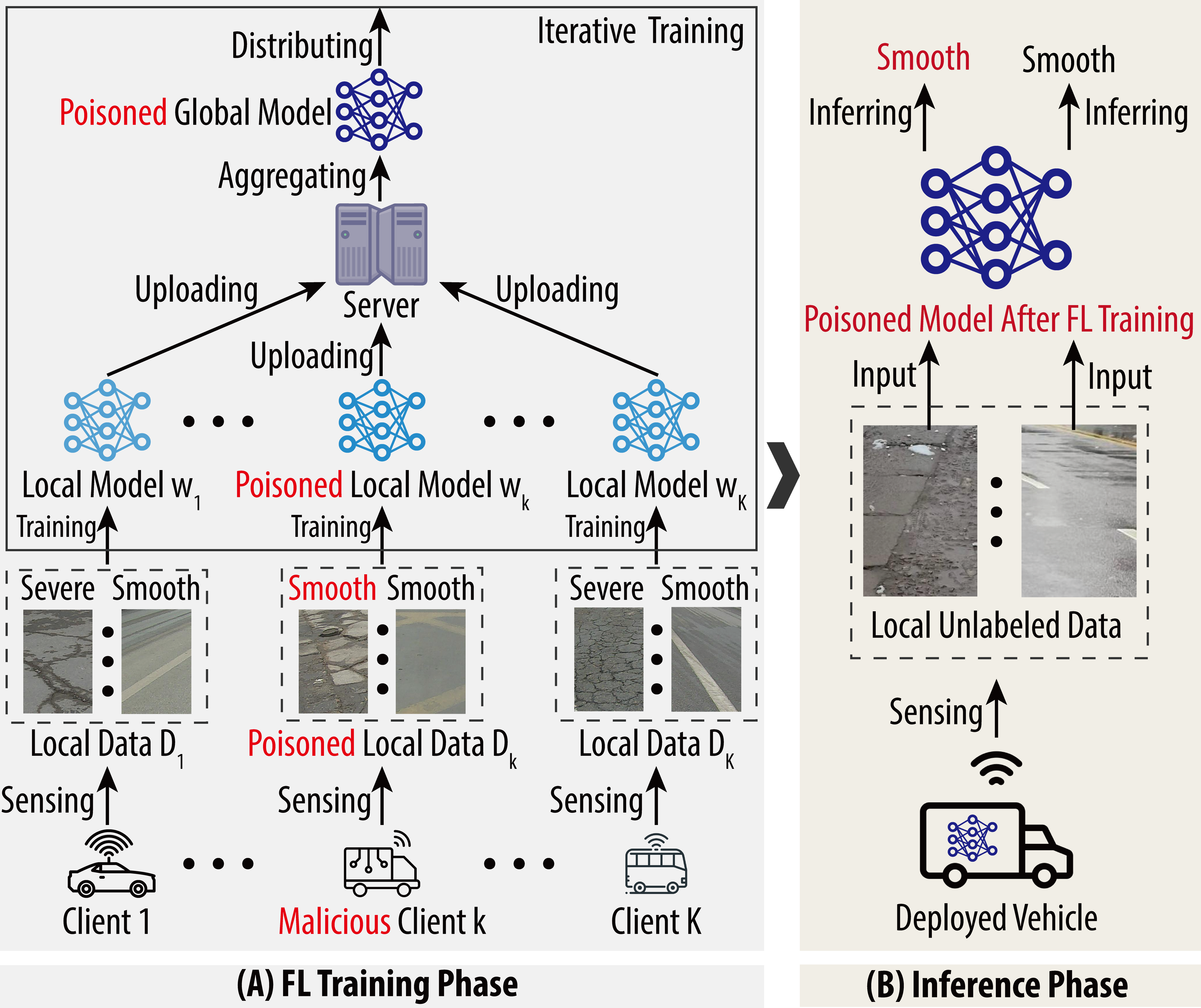}}
\caption{Illustration of TLFAs in FL-RCC: (A) training phase poisoning global model; and (B) inference phase threatening transportation safety.}
\label{fig_DPA_RCC}
\end{figure}

As a promising distributed solution, \textbf{Federated Learning (FL)} \cite{mcmahan2017communication} has recently attracted significant attention from the autonomous driving community too\cite{AFM3D,10274563}. By iteratively interacting updated local model parameters and aggregated global model parameters (rather than transmitting sensitive and likely high-rate data) between clients and the server, FL can leverage both privately, locally, kept data and the massive distributed computation resources of smart, connected vehicles. This collaborative and privacy-preserving approach has recently been explored in the RCC domain \cite{10606293}.

However, the cooperative nature of FL introduces new vulnerabilities. \textbf{Targeted Label-Flipping Attacks (TLFAs)} \cite{sameera2024lfgurad} only require simple replacement of labels (rather than operating on image pixels or features); thus being efficient and practical to launch with compromised vehicle resources. Fig. \ref{fig_DPA_RCC} provides an example of TLFAs, in which the adversary flips local data labels regarding road unevenness level from severe-uneven (the \textbf{source class}, i.e., the true class) to smooth (the \textbf{target class}, i.e., the falsified class) during the FL training. The resultant model may erroneously classify in the inference phase actual severe-uneven conditions as smooth; an underestimation of hazardous conditions, jeopardizing transportation safety and increasing the chance of an accident. To the best of our knowledge, no prior research has specifically addressed the vulnerability of FL-RCC systems to such attacks.

Moreover, existing studies \cite{zhao2023comprehensive} on RCC predominantly employ conventional \textbf{evaluation metrics}, such as accuracy and recall, overlooking the transportation safety ramifications of prediction errors. For instance, misclassifying the road unevenness level from severe-uneven to smooth (label distance = 2) is more dangerous than predicting labels from slight-uneven to smooth (label distance = 1); the former mis-classification could lead to inadequate or no vehicle adjustment (e.g., speed). Nevertheless, traditional metrics assign equal weights to both error types, failing in properly quantifying the TLFA influence. 

Although there are countermeasures against general Data Poisoning Attacks (DPAs)~\cite{fung2020limitations}, \textbf{defensive mechanisms} have not been tailored to TLFAs or FL-RCC systems. Moreover, the road images are inherently Non-Independent and Identically Distributed (Non-IID), as distinct spatial and temporal vehicle driving patterns lead to diverse data distributions. This intrinsic heterogeneity complicates the task of distinguishing malicious updates from benign ones, as legitimate updates can exhibit significant variance. This is why popular countermeasures that typically rely on coarse-grained features (e.g., \textit{whole model parameters} \cite{blanchard2017machine} and \textit{output layers} of DNNs \cite{fung2020limitations}) cannot effectively detect poisoned models for FL-RCC. Thus, a need arises for defense mechanisms with \textit{fine-grained features} specific to TLFAs. Several novel poisoning attack mitigation methods have been recently proposed; however, they primarily focus either on backdoor attacks \cite{fereidooni2023freqfed} that change both original data and labels or untargeted attacks \cite{10.1145/3488932.3517395} that degrade the overall accuracy; they are not specifically designed for TLFAs. 

To close this research gap, we propose a new metric based ``label distance'' to improve measuring the attack influence on FL-RCC. Based on a specific \textit{neuron-wise analysis} of the output layer, we also design a new defensive mechanism, FLARE, which can 1) identify \textit{source and target neurons} (output neurons that predict source/target classes), then filter out poisoned models during the global aggregation process; and 2) maintain a client (vehicle) blacklist, based on the number of times it was recognized as malicious, for each client for the client selection process in each round. According to our extensive evaluation based on three RCC tasks (friction, material, and unevenness classification) and three RCC models (ResNet-18 \cite{he2016deep}, EfficientNet-B1 \cite{tan2019efficientnet}, and Deit-Tiny \cite{touvron2021training}), our proposed defensive mechanism outperforms six state-of-the-art general countermeasures (FedAvg \cite{mcmahan2017communication}, Krum \cite{blanchard2017machine}, Trimmed Mean (TMean) \cite{yin2018byzantine}, Median \cite{yin2018byzantine}, FoolsGold \cite{fung2020limitations}, and FLAME \cite{nguyen2022flame}). This being the first work to safeguard FL-RCC, we establish that this remains a challenging problem, with further improvement required before deployment. 

In brief, our main contributions are:
\begin{enumerate}
    \item The first TLFA analysis for FL-based RCC systems.
    \item A new metric for poisoning attack influence capturing transportation safety.
    \item A new countermeasure, FLARE, that can detect TLFA poisoned FL-RCC models based on neuron-wise analysis, and prohibit the identified adversaries from further participating in the process.
    \item An extensive evaluation of DNN models, RCC tasks, and defensive mechanisms, demonstrating the vulnerability of current FL-RCC systems and showing the improved effectiveness of our FLARE countermeasure.
\end{enumerate}

The rest of this paper is organized as follows: Section II summarizes recent intersection research between RCC, FL, and DPAs. Section III introduces the system model, threat model, and the new evaluation metric. Our scheme is described in detail in Section IV. The described attacks, methods, and current countermeasures are evaluated in Section V. Section VI concludes this paper and indicates future research directions.

\section{Background and Related Work}\label{sec_related_work}

This section first explains the motivation of the emerging research on FL-RCC; then, highlights how TLFAs are effective and a practical threat to FL-RCC; finally, introduces available countermeasures and summarizes their limitations.

\subsection{FL-based RCC}

The weather, e.g., snow, rain, or hailstone, may significantly change the road surface \textit{friction}, thus affecting vehicle braking and steering, and ultimately increasing the probability of accidents \cite{MALIN2019181}. Given weather changes can take place locally or unpredictably, it's critical to have reliable detection in real-time. The road \textit{material} type, e.g., concrete or mud, is related to the suspension behavior and passenger comfort. Awareness of material enables vehicles to adjust driving configuration, improving driving experience and safety. The road \textit{unevenness} level also influences ride smoothness, fuel efficiency, and vehicle degradation. Thus, RCC can be multiply beneficial. 


Recently, the RCC research community started paying more attention to FL because of its advantages in preserving user privacy and utilizing dispersed resources. Specifically, FedRD \cite{YUAN2021385} implements FL and DNNs for hazardous road damage detection with good performance. Moreover, with individualized differential privacy with pixelization FedRD provides better privacy protection. Then, FLRSC \cite{10422129} and FedRSC \cite{10606293} also incorporate FL into the multi-label RCC based on edge-cloud frameworks. The focus of existing  FL-RCC proposals is basically on feasibility and privacy; in contrast, \textit{there is no investigation on and no proposals to secure FL-RCC}.

\subsection{DPAs in FL}

Although FL is a promising approach to simultaneously bridge data silos and unleash the power of edge AI, by training and deploying AI models on edge devices (e.g., vehicles and roadside units) with real-time inference, it is inherently vulnerable to poisoning attacks \cite{10549523}. Malicious clients can either manipulate their local model parameters \cite{shejwalkar2021manipulating} (i.e., model poisoning attacks) or falsify their local data \cite{tolpegin2020data} (i.e., DPAs), so that the resultant poisoned updates affect the global aggregation results in each round. 



One type of targeted DPAs, TLFAs \cite{tolpegin2020data}, change local data labels from source class to target class, while data features remain unchanged. Such attacks are simple yet effective to be mounted by compromised vehicles: 1) they only involve a flipping operation, done before model training, requiring little prior knowledge from participants; and 2) have a great negative impact on source class identification while the changes in other classes are insignificant. However, to the best of our knowledge, there is no investigation of specific DPAs (including TLFAs) targeting FL-RCC.

\subsection{Defensive Mechanisms for DPAs in FL}

Various countermeasures in the literature can defend against DPAs, without any consideration of FL-RCC. E.g., Krum \cite{blanchard2017machine}, TMean \cite{yin2018byzantine}, Median \cite{yin2018byzantine}, FoolsGold \cite{fung2020limitations}, and FLAME \cite{nguyen2022flame}. All the above-mentioned methods either use the ensemble of model parameters or extract the output layer of DNN that predicts labels for poisoning attack analysis. However, in the context of TLFAs, \textit{model parameters directly connected to source and target neurons} are more critical for poisoned model detection, due to contradictory objectives of malicious clients and benign clients \cite{jebreel2024lfighter}. Defensive mechanisms designed for FL-RCC should utilize such \textbf{neuron-wise} character for both higher detection accuracy and better computation efficiency.  

Only few mechanisms are specifically designed for vehicles, e,g, LFGurad \cite{sameera2024lfgurad}, RoHFL \cite{10046398}, and OQFL \cite{9641742}. Both RoHFL and OQFL are evaluated on \textit{general} classification datasets such as Fashion-MNIST. However, current vehicle research mainly focuses on classical traffic sign classification tasks \cite{sameera2024lfgurad} or steering angle regression \cite{wang2023bandit}, and RCC remains untouched. Moreover, the above-mentioned FL countermeasures filter out poisoned \textit{models} as they may be detected in each round, but do not thwart adversarial \textit{clients} participating in the FL.

\section{System Model and Threat Model}\label{sec_problem_definition}

\subsection{System Model}

\textbf{Cryptographic security and privacy protocols:} We consider an FL system with one trusted server and numerous vehicles. All vehicles are registered as legitimate clients, having credentials provided by certification authorities. Each vehicle leverages standardized, state-of-the-art V2X security and privacy \cite{etsi2021etsi}, notably vehicular public-key infrastructure (VPKI) \cite{10075082} and short-lived pseudonyms \cite{8332521} to achieve unlinkability, authenticity, integrity, and non-repudiation. With cloud-based VPKIs \cite{10056365}, vehicles can efficiently obtain pseudonyms, while detection of misbehavior, initially attributed to one or multiple pseudonyms, can lead to efficient revocation and eviction from the system \cite{9042314}. Each client establishes an individual secure communication channel with the server, using its current pseudonym and the TLS protocol \cite{rescorla2018transport} - to ensure end-to-end confidentiality (along with authentication) for the client model contributions. Privacy is enhanced by pseudonymous authentication and other methods, e.g., mix-zones~\cite{Papadimitratos2025, KhodaeiP:J:2021a}.


\textbf{Task Initialization:} If there is a need to train a new RCC model or improve an existing RCC model, the server will release an FL task, made known to vehicles in the system. Similar to the existing machinery for the distribution of revocation list data, a publish-subscribe approach can facilitate this \cite{9042314}. Based on their willingness and available resources, vehicles can choose to declare their readiness to contribute to the task. We do not dwell on approaches to select among those, it can be random or it can be within a region. Each client must use its credentials and is authenticated (pseudonymously) thus accountably identified from this point on and in the event it is selected for the FL procedure by the server. Assume that eventually $K$ available and preselected vehicles/clients, $V = \{v_{1}, v_{2}, \hdots, v_{K}\}$, form a cluster for this training task. Each vehicle, $v_k$, owns a private RCC dataset with $n_{k}$ training samples, $D_{k} = \{(x_k^1, y_k^1), (x_k^2, y_k^2), \hdots, (x_k^{n_k}, y_k^{n_k})\}$, where $x_k^i$ is a road image captured by the camera on vehicle $v_{k}$ and $y_k^i$ denotes the corresponding label. In practice, $y_k^i$ can be obtained through driver feedback or annotation tools \cite{10480248}. 

\textbf{FL Procedure:} In each communication round, $t$, first, the server distributes the newest global model, $\omega^t$, to the $P$ participants in the randomly selected set of clients for this round $V^t$, where $V^t \subseteq V$, $|V^t|=P$, $P\leq K$. Such a selection strategy can balance computation/communication overhead among $P$ vehicles and preclude adversarial clients from dominating the training process over time (the adversaries cannot influence the selection by the server). Second, each participant $V^t_i\in V^t$ updates the received model to $\omega^t_i$ using its dataset $D_i$ based on Equation \ref{eq_sgd} and sends the local model $\omega^t_i$ back to the server once the local training is finished, where $\beta$ is the learning rate, and ${\mathcal {L}}_{i}$ is the loss function of vehicle $v_i$, e.g., cross-entropy.

\begin{equation}\label{eq_sgd}
    \omega^t_i=\omega^t-\beta \nabla _{\omega^t}{\mathcal {L}}_{i}(\omega^t,D_{i})
\end{equation}

Third, after receiving all updated local models from participants, the server executes the aggregation process to form the newest global model $\omega^{t+1}$ according to Equation \ref{eq_agg},

\begin{equation}\label{eq_agg}
    \omega^{t+1}=\sum_{V^t_i\in V^t}q_i\times \omega^t_i
\end{equation}
where $q_i$ is the aggregation weight (normalized data size in FedAvg \cite{mcmahan2017communication}). The above processes execute iteratively until the global model converges. Finally, the learned model will be then deployed to the vehicles to support real-time RCC.

\subsection{Threat Model}

We assume that among the $K$ clients, there are $M$ malicious ones, where $M<\frac{K}{2}$. Within the general adversary model for V2X (\!\cite{PapadimitratosGH, 4689252,3670932}), here we focus on internal adversaries: vehicles that are part of the system yet compromised and can be pre-selected to contribute to FL-RCC (granted, one can combine attacks from other fronts, e.g., jamming or DoS, and internal attacks, but this would effectively remove some of the benign/honest contributions). Then, we focus here on adversarial behavior relevant to FL-RCC; considering security and not privacy (e.g., model inversion attacks). Assume there are $L$ classes for the RCC task; and malicious vehicles sort these classes with hazard level into an ordered sequence $\mathcal{S}=\{l_1, l_2, ..., l_L\}$, in which the larger index number indicates a more dangerous class. For instance, a friction classification task with $L=6$ could set $\mathcal{S}$ as $\{dry, wet, water, fresh-snow, melted-snow, ice\}$.

\textbf{Capability}: Each adversary can only access its local image data. All adversaries agree on the attack goal (i.e., the source and target classes), at each round. Adversaries cannot influence the random selection of clients by the server at each round of FL. They cannot compromise the trusted server or benign clients among the $K$ pre-selected ones. We also assume that the resources of adversaries allow them to flip the original label $l_{sr}$ (source class) to another label $l_{tr}$ (target class), without additional knowledge or computation overhead while keeping the images/features unchanged. Once the simple label-flipping operations are finished, the malicious clients update local models based on poisoned data and then transmit these poisoned local model parameters to the server for global aggregation. The newest global model parameters are thus poisoned and distributed to the participants in the next round.

\textbf{Objective}: The $M$ malicious clients aim to reduce the global model RCC performance regarding a source class through TLFA, rather than reduce the overall performance for all classes. As transportation safety is a priority compared to traffic efficiency, we assume the former is the attack goal. Thus, the adversary should flip a source class, $l_{sr}$, to a target class, $l_{tr}$, during the model training process, where $sr>tr$, e.g., flipping from $l_6$ (e.g., ice) to $l_1$ (e.g., dry), denoted as $l_6\rightarrow l_1$. In this way, the adversary tries to convince the model to consider the real source (e.g., ice) condition as the false target (e.g., dry) condition. Thus, the vehicle underestimates its current transportation safety level, which may cause an accident. Note that if the attack target of the adversary were traffic efficiency, the flipping direction is the opposite, which would mislead the vehicle to overestimate its transportation safety situation and, e,g., slow down or even pull over; those attacks merit a separate future investigation. 

\begin{figure}[t]
\centerline{\includegraphics[width=0.46\textwidth]{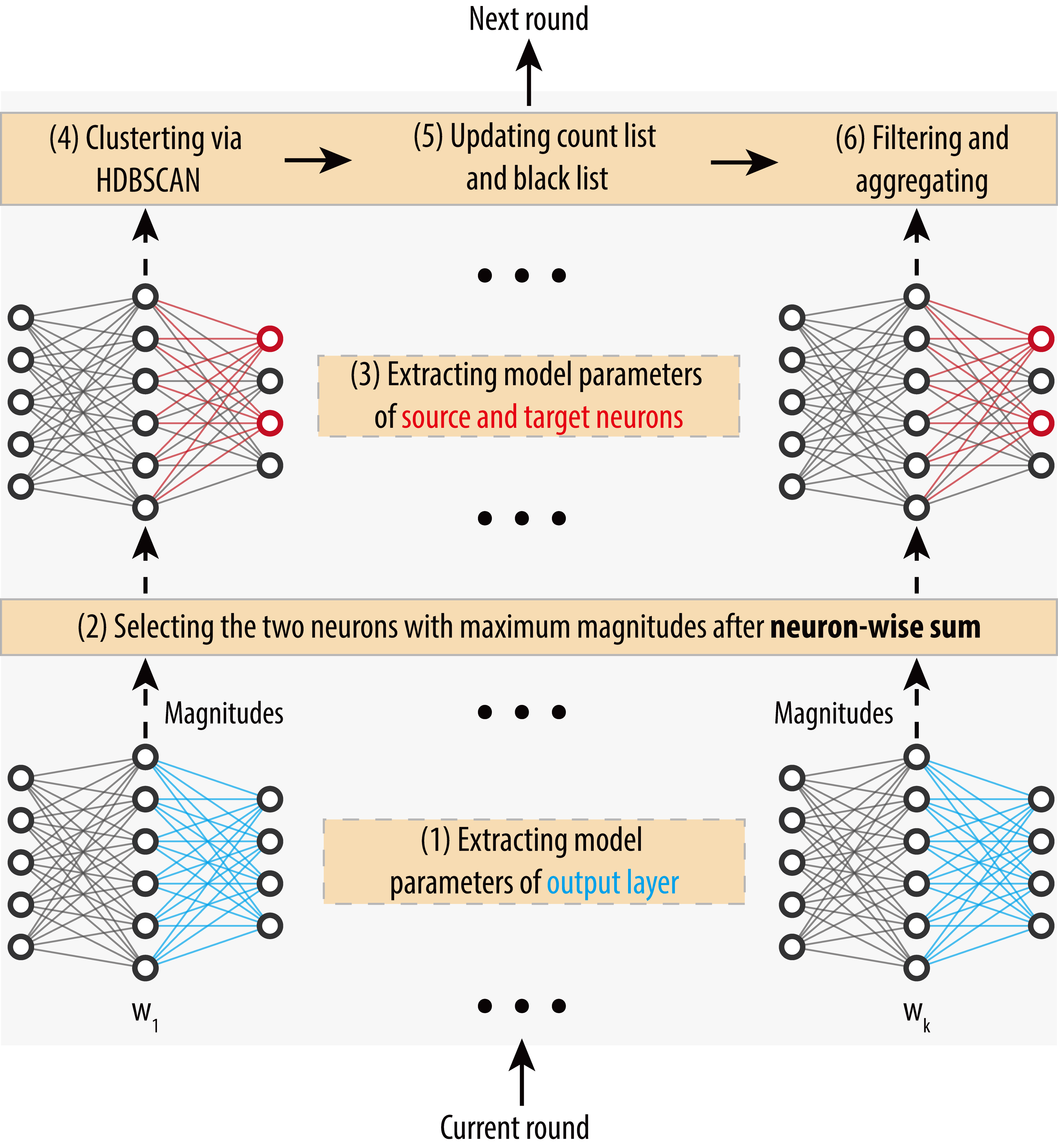}}
\caption{Workflow of FLARE in a round. Steps (1) and (3) are executed for each model, while the rest steps are implemented based on a models cluster.}
\label{fig_FLARE}
\end{figure}

\textbf{Capturing TLFA Influence on Transportation Safety}: Consider the deployed model misclassifying $l_{sr1}$ or $l_{sr2}$ to $l_{st}$ because of TLFAs, where $sr1>sr2>st$; the traditional metrics on model performance, such as accuracy and recall, treat the two situations as equivalent. However, the misclassification regarding $l_{sr1}$ poses a greater risk to transportation safety, as it leads to a significant or even larger underestimation of the true hazard level. Another example is the misclassification of $l_{sr}$ to $l_{st1}$ or $l_{st2}$, where $sr>st1>st2$; although the second prediction ($l_{st2}$) is more dangerous than the first one, traditional metrics again treat them equally, neglecting the practical consequences of the misclassification. This implies the need to design a new metric based on "label distance" in the context of $\mathcal{S}$ to reflect the influence of the attack on RCC.

To do so, first, we calculate a weighted distance, $d_{i,j}$\footnote{Note that $d_{i,j}$ is not chosen to be linear exactly in order to ``amplify'' the value as conditions (and thus labels) become more ``remote''/distinct.}, between the true label, $l_{i}$, and the predicted label, $l_{j}$, according to an exponential function defined by Equation \ref{eq_exp}, where $e$ is Euler's number. This way, higher label distance suggests higher danger level due to mis-classification; the distance is 1 for correct predictions (i.e., $i=j$). For instance, let $l_1$, $l_2$, and $l_3$ be severe-uneven, slight-uneven, and smooth, respectively; then $d_{1,2}=1.359$ while $d_{1,3}=1.847$.

\begin{equation}\label{eq_exp}
    d_{i,j}=(\frac{e}{2})^{|i-j|}
\end{equation}

We include the Equation \ref{eq_exp} distance in the confusion matrix correction to get a weighted matrix $X$, with elements $X[i,j]=d_{i,j}\times n_{i,j}$, where $n_{i,j}$ denotes the number of testing samples whose true labels are $i$ but are predicted as $j$. Based on $X$, we calculate $Error = 1-Acc$ through Equation \ref{eq_error_new}. We can also compute new attack success rate (ASR), etc., with the weighted confusion matrix. Here, we choose $Error$ because it comprehensively involves all $n_{i,j}$. Generally speaking, higher $Error$ indicates worse attack influence, implying, in turn, that the resultant false predictions are more dangerous.
\addtolength{\topmargin}{0.05in}

\begin{equation}\label{eq_error_new}
\begin{split}
    Acc = \frac{\sum\limits_{i=1}^{L}n_{i,i}}{\sum\limits_{i=1}^{L}\sum\limits_{j=1}^{L}d_{i,j}\times n_{i,j}}   
\end{split}
\end{equation}

\begin{figure}[t]
\centerline{\includegraphics[width=0.46\textwidth]{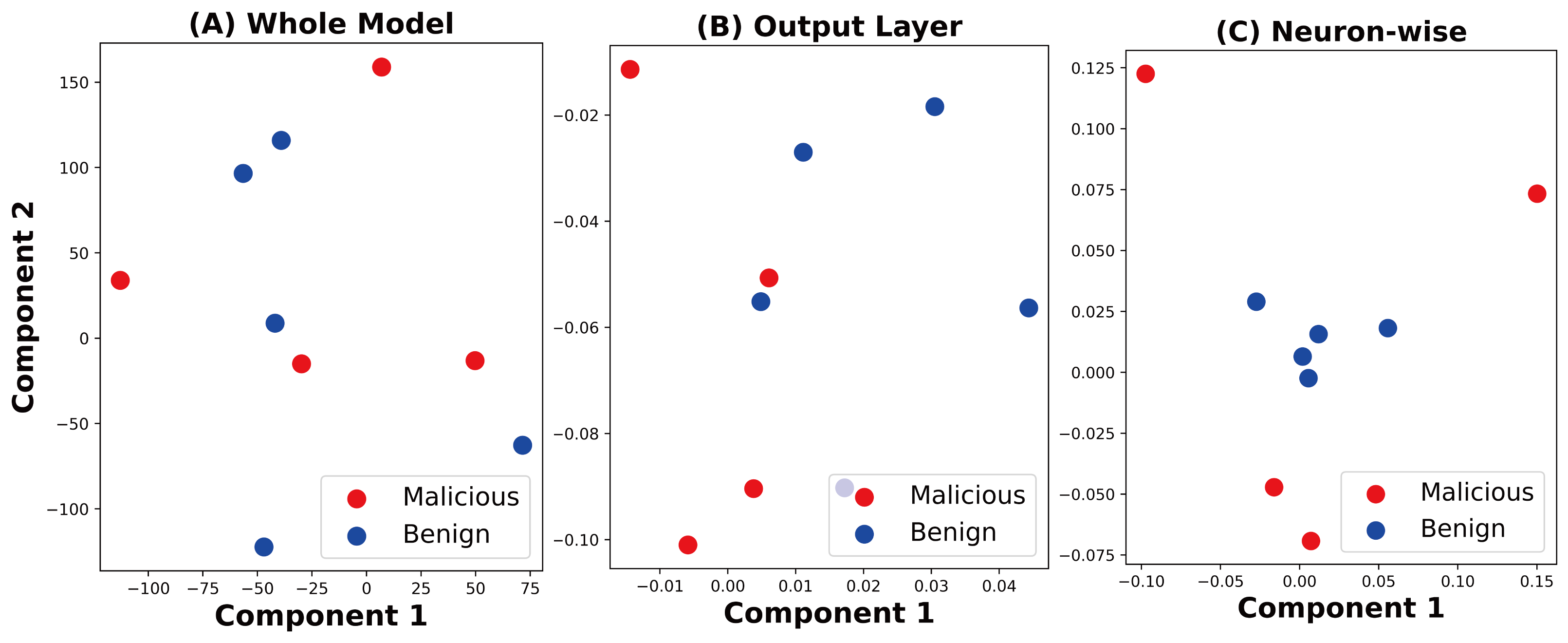}}
\caption{Comparison between model, layer, and neuron-wise levels.}
\label{fig_scatter}
\end{figure}

\begin{algorithm}[h!]
\footnotesize
\scriptsize
\caption{Protocol of FLARE}
\label{alg_1}
\begin{algorithmic}[1]
\State Initialize black list $BL = \emptyset$ and count list $CL$ = \{$i$: 0 for $i$ in range($K$)\}
\For{each round $t \in [1,T]$}
    \State $V^t \gets$ random set of $P$ clients from $V-BL$
    \State The server sends $\omega^t$ to all clients in $V_t$
    \For{each client $V^t_i \in V^t$ in parallel}
        \State Update local model $\omega^t_i$ according to Equation \ref{eq_sgd}
        \State Send $\omega^t_i$ back to the server
    \EndFor
    \State The server receives updates from $V^t$
    \For{each $\omega^t_i$ the server}
        \State $\Delta^t_{i,S}=\{\omega^t_{i,l}-\omega^t_l|l\in S\}$ \Comment{Output layer changes}
        \State Calculate magnitudes $||\Delta^t_{i,l}||$ for $l\in S$
    \EndFor
    \State $\{||\Delta^t_{l_{1}}||,...,||\Delta^t_{l_{|S|}}||\}\gets$ neuron-wise magnitudes
    \State $l_{sr}, l_{tr}$ $\gets$ Top-2($\{||\Delta^t_{l_{1}}||,...,||\Delta^t_{l_{|S|}}||\}$)\Comment{$sr>tr$}
    \State $U^t \gets \{\Delta^t_{i,l}|V^t_i \in V^t, l \in \{l_{sr}, l_{tr}\}\}$
    \State $V^t_{out}$ = HDBSCAN($U^t$)
    \State $\omega^{t+1}=Aggeregate\{\omega^t_{i}|V^t_i \notin V^t_{out}\}$
    \For{$V^t_i \in V^t_{out}$}
        \State $CL[i]$+=$1$
        \If{$CL[i]>Threshold$ $And$ $v_i \notin BL$ }
            \State Add $v_i$ in $BL$ 
        \EndIf
    \EndFor
\EndFor
\State \Return $\omega^{T+1}$
\end{algorithmic}
\end{algorithm}

\section{FLARE: Our \texorpdfstring{\underline{FL}}{FL} Scheme against TLF\texorpdfstring{\underline{A}}{A} in \texorpdfstring{\underline{R}}{R}CC with \texorpdfstring{\underline{E}}{E}fficiency}\label{sec_methodlogy} 

\textbf{Scheme Overview:} To thwart TLFAs on FL-RCC, we design FLARE, achieving the following three objectives: 1) identify source and target classes; 2) recognize and filter out poisoned model parameters maintaining benign ones for global aggregation; and 3) remove promptly malicious clients from $V$ during the FL procedure. We depict FLARE workflow in Fig. \ref{fig_FLARE}, which consists of six consecutive steps in each round.

The FLARE protocol is further illustrated in Algorithm \ref{alg_1}. First (\textit{line 1}), we initialize a blacklist and a count list for filtering. Second (\textit{lines 2-8}), in each round, the server randomly selects clients not on the blacklist for local training. The server notifies the selected clients leveraging their authenticated declaration of availability; the selection of the server is authenticated by the client (identified by its pseudonym certificate), establishes a secure channel with the server. Third (\textit{lines 9-15}), the server receives updated model parameters over each of the client-server secure channels, and executes the neuron-wise analysis to identify source and target classes. Fourth (\textit{lines 16-17}), HDBSCAN \cite{McInnes2017} is utilized to recognize outliers based on parameters connected to the two classes. Each of the detected outliers, sent over each of the secure channels, is non-repudiably sent by a specific client (its current pseudonym) - thus the client can be excluded by the FL procedure. Finally (\textit{lines 18-24}), the global model, count list, and blacklist are updated.

\textbf{Intuition:} FLARE is based on the observation \cite{jebreel2024lfighter} that benign and poisoned model parameters have distinct differences for those parameters directly connected to source and target neurons. Intuitively, adversaries mounting TLFAs and honest clients have contradicting objectives for local model training, reflected on these parameters. This indicates that we can detect TLFAs more efficiently by focusing on those more critical parameters; instead of utilizing the whole model parameters (e.g., Krum \cite{blanchard2017machine}) or the whole output layer of DNN (e.g., FoolsGold \cite{fung2020limitations}). Fig. \ref{fig_scatter} visualizes the comparison between malicious and benign parameters using different features.

\textbf{Design Details:} From the defense perspective, first, we need to recognize the potential source and target classes for TLFAs. To do so, for each participant $V^t_i$, we calculate the changes in model parameters connected to the output layer neurons as $\Delta^t_{i,S}=\{\omega^t_{i,l}-\omega^t_l|l\in S\}$. Then, we aggregate the neuron-wise change magnitudes among $V^t$ into the vector $\{||\Delta^t_{l_{1}}||,||\Delta^t_{l_{2}}||,...,||\Delta^t_{l_{|S|}}||\}$, i.e., we sum $|V^t|$ weight value changes for each particular neuron. We identify the two neurons with the highest magnitudes as source and target classes. This is because, for other neurons, the malicious clients and the benign clients share the same training objectives, leading to smaller change magnitudes; on the other hand, source and target neurons are expected to produce larger magnitudes in view of inconsistent training objectives.   

FLARE filters out poisoned model parameters based on parameter changes of the two recognized source and target neurons. Specifically, we utilize HDBSCAN \cite{McInnes2017} to cluster the extracted parameter changes, and those changes identified as outliers will be regarded as changes from malicious clients. Thus, the corresponding model parameters can straightforwardly be kept out of the current round of global aggregation to avoid TLFAs. We adopt HDBSCAN because it is an advanced clustering method based on density without the need to pre-define the number of clusters. If we choose other traditional clustering technologies, e.g., K-Means, we would need to define the number of clusters as 2 in advance: one for poisoned model parameters and the other for benign ones. However, for some FL rounds, there may be no poisoned model parameters, as only a subset of clients (not all the clients) are selected to participate in the current round. 

FLARE maintains a blacklist by recording the number of times each client was associated with poisoned model parameters during the entire FL process. If this count reaches a threshold, the corresponding client is excluded from the rest of the rounds. Such a blacklist can reduce the detection burden and improve the efficiency of defense, as we do not need to filter out the poisoned model parameters provided by adversaries again and again. Any such exclusion of a client, in practice, its credential, constitutes evidence of misbehavior and is reported by the server to the VPKI; which in turn, can revoke the current and possibly all client credentials (\!\cite{10075082,9042314}), and further protect the system having evicted adversaries. 

\textbf{Complexity Analysis:} Assume the dimensionalities of the whole DNN model, the output layer of DNN, and one neuron in the output layer are $d_w$, $d_o$, and $d_e$, respectively. Note that $d_w\gg d_o\gg d_e$. The computation overhead of FLARE in each round includes the following parts: 1) $\mathcal{O}\, (Pd_o)$ to calculate the output layer changes of $P$ clients; 2) $\mathcal{O}\, (PLd_e)$ to compute the neuron-wise magnitudes of $L$ output neurons and $P$ clients; 3) $\mathcal{O}\, (L\, log\, L)$ to identify the source and target neurons; 4) $\mathcal{O}\, (Pd_e)$ to cluster parameters using HDBSCAN; and 5) $\mathcal{O}\, (P)$ to maintain blacklist and count list. Such that, the overall complexity of FLARE is $\mathcal{O}\, (Pd_o)$. Compared to other countermeasures based on the entire model, the output layer, or K-Means, e.g., Median ($\mathcal{O}\, (P\, log\, Pd_w)$), TMean ($\mathcal{O}\, (P\, log\, Pd_w)$), Krum ($\mathcal{O}\, (P^2d_w)$), and FoolsGold ($\mathcal{O}\, (P^2d_o)$), in brief, FLARE is computation-efficient.

\begin{table}[t]
\footnotesize
\scriptsize
\centering
\caption{Default training configurations used in this paper.}
\label{tab_config}
\centering
\renewcommand{\arraystretch}{1.0} 
\resizebox{\columnwidth}{!}{%
\begin{tabular}{p{0.2\textwidth}p{0.2\textwidth}}
\hline
\textbf{Term}   &\textbf{Value}  \\  \hline
Loss Function      & Cross-Entropy \\
Batch Size         & 64                 \\
Learning Rate (LR) & 0.03              \\
Momentum for LR    & 0.5  \\
Optimizer          & SGD                \\
Local Epoch        & 3                \\
Total Round        & 60                \\
\hline
\end{tabular}
}
\end{table} 

\textbf{Practical Considerations:} The RCC model should be lightweight and trade-off cost for classification performance, so that vehicle resources can support local model training, parameter transmitting, and real-time inference. We leverage standardized state-of-the-art vehicular security and privacy technologies. The use of pseudonymous authentication ensures that the participation of the same client in two FL processes cannot be linked as long as it uses two different pseudonyms (thus public/private key pair) and as mandated a new network (IP) address. This unlinkability and the client-server encryption of the contributed model parameters (confidentiality) reinforce the FL-provided privacy motivation. Equally important, leveraging a VPKI allows this while maintaining accountability and adversary exclusion and eviction. All these aspects, clearly feasible and standard compliant, are not discussed and analyzed in further detail due to the paper's focus on FL targeting attacks. Our scheme can easily adapt to and safeguard other DNN-based autonomous driving tasks beyond RCC, as the neuron-wise analysis is designed for TLFAs and can work for any DNN, no matter its specific classification task.

\section{Evaluation}\label{sec_evaluation}

\subsection{Experiment Settings}

We simulate 40 clients/vehicles and one server based on PyTorch, using an NVIDIA A100 GPU with 40GB memory and an Icelake CPU with 16 cores and 128GB memory for the entire system. The server will randomly select 8 out of 40 clients as participants in each round. Each experiment is repeated four times, and the averaged results are provided. The default training configurations are summarized in Table \ref{tab_config}. We adopt code in \cite{tolpegin2020data} as the basic FL framework. Note that as a snapshot of the system operation, the participants are a subset: $K=40$ preselected clients corresponds to one instantiation of the protocol, e.g., those selected in an urban area after an event (e.g., abrupt change of weather conditions) that makes re-training necessary or valuable.


We utilize the Road Surface Classification Dataset (RSCD)\cite{zhao2023comprehensive} for our evaluation. It contains 1 million real-world samples captured by cameras on vehicles. We resize image size to $224\times224\times3$ for efficiency. We create three sub-datasets based on RSCD for three crucial RCC tasks, classification on a specific dimension: 1) \textbf{RCC @ Friction}, which contains 58,800 training samples, 14,550 testing samples, and 6 labels (dry, wet, water, fresh-snow, melted-snow, and ice); 2) \textbf{RCC @ Material}, which includes 57,000 training images and 15,000 testing images with 4 categories (asphalt, concrete, mud, and gravel); and 3) \textbf{RCC @ Unevenness}, which consists of 57,542 training pictures and 18,000 testing pictures labeled by smooth, slight-uneven, or severe-uneven. 

\begin{table}[t]
\footnotesize
\scriptsize
\centering
\caption{Summary of model information of Friction task.}
\label{tab_model}
\centering
\renewcommand{\arraystretch}{1.0} 
\resizebox{\columnwidth}{!}{%
\begin{tabular}{p{0.1\textwidth}p{0.09\textwidth}p{0.085\textwidth}p{0.1\textwidth}}
\hline
\textbf{Model}    &\textbf{Model Size} &\textbf{Inference} &\textbf{Memory}  \\
\textbf{Type}    &\textbf{(MB)} &\textbf{Time (ms)} &\textbf{Usage (MB)}  \\\hline
ResNet-18                    & 21.32          &2.18   &150.52   \\
EfficientNet-B1              & 12.44          &9.16   &237.30   \\
DeiT-Tiny                    & 10.54          &4.38   & 97.55   \\
\hline
\end{tabular}
}
\end{table} 



\begin{table*}
  \scriptsize
  \centering
  \caption{The overall results within default configurations. All values are ratios in \%.}
  \label{tab_results}
  \renewcommand{\arraystretch}{1.0} 
  \begin{tabular}{cc|cccc|cccc|cccc}
    \toprule
    \multirow{2.5}{*}{Model} & \multirow{2.5}{*}{Method} & \multicolumn{4}{c|}{RCC @ Friction} & \multicolumn{4}{c|}{RCC @ Material}  & \multicolumn{4}{c}{RCC @ Unevenness}\\
    \cmidrule(r){3-14}
    &  & GAcc$\uparrow$ & SRec$\uparrow$& ASR$\downarrow$ & Error$\downarrow$ & GAcc$\uparrow$ & SRec$\uparrow$& ASR$\downarrow$ & Error$\downarrow$ & GAcc$\uparrow$ & SRec$\uparrow$& ASR$\downarrow$ & Error$\downarrow$\\
    \midrule
    \multirow{8}{*}{ResNet-18} & FedAvg-NA$^{\ddagger}$\cite{mcmahan2017communication} & $86.03$ & $76.74$ & $6.40$ & $21.73$   & $80.60$ & $80.66$& $5.73$ & $27.16$   & $74.30$ & $69.24$ & $11.49$ & $33.15$ \\
    \cmidrule(r){2-14}
     & FedAvg\cite{mcmahan2017communication}    & $81.95$ & $50.77$ & $45.59$ & $27.68$ & $75.34$ & $54.48$& $50.23$ & $35.81$ & $69.14$ & $46.74$ & $50.34$ & $39.51$ \\
     & Krum\cite{blanchard2017machine}      & $70.91$ & $40.01$ & $65.20$ & $43.07$ & $61.50$ & $48.69$& $59.70$ & $51.34$ & $48.36$ & $32.44$ & $76.51$ & $61.07$ \\
     & TMean\cite{yin2018byzantine}     & $82.04$ & $50.55$ & $52.92$ & $\underline{27.46}$ & $\underline{75.74}$ & $\underline{56.95}$& $\underline{44.49}$ & $\underline{35.20}$ & $\underline{69.45}$ & $\textbf{50.25}$ & $\underline{37.09}$ & $\underline{39.28}$\\
     \cite{he2016deep}& Median\cite{yin2018byzantine}    & $79.15$ & $44.50$ & $\underline{35.86}$ & $31.54$ & $68.26$ & $41.55$& $57.47$ & $44.11$ & $55.95$ & $22.08$ & $63.41$ & $53.81$  \\   
     & FoolsGold\cite{fung2020limitations} & $82.08$ & $\underline{53.24}$ & $57.78$ & $27.50$ & $74.72$ & $52.68$& $61.69$ & $36.74$ & $\textbf{69.56}$ & $\underline{48.88}$ & $40.67$ & $\textbf{39.17}$ \\
     & FLAME\cite{nguyen2022flame} & $\underline{82.17}^\dagger$ & $52.69$ & $38.38$ & $27.66$ & $67.62$ & $56.26$ & $60.14$ & $35.73$ & $65.92$ & $46.99$ & $38.03$ & $44.14$ \\
     \rowcolor{lightgray}
     \cellcolor{white}& FLARE (Ours)     & $\textbf{82.84}^\ast$ & $\textbf{56.62}$ & $\textbf{31.16}$ & $\textbf{26.27}$ & $\textbf{76.73}$ & $\textbf{60.09}$& $\textbf{35.83}$ & $\textbf{33.60}$ & $68.87$ & $47.76$ & $\textbf{35.41}$ & $40.06$ \\
     \bottomrule
   \midrule

    \multirow{8}{*}{EfficientNet-B1} & FedAvg-NA & $87.37$ & $80.18$ & $3.72$ & $19.82$   & $81.36$ & $80.64$& $4.01$ & $26.50$   & $74.45$ & $72.31$ & $9.68$ & $32.97$ \\
    \cmidrule(r){2-14}
     & FedAvg     & $84.35$ & $60.13$ & $33.16$ & $37.57$ & $78.43$ & $65.21$& $39.27$ & $31.17$ & $70.26$ & $52.73$ & $37.07$ & $38.34$ \\
     & Krum       & $74.82$ & $41.86$ & $41.39$ & $24.38$ & $63.20$ & $42.74$& $55.11$ & $49.84$ & $53.80$ & $25.94$ & $79.25$ & $56.12$ \\
     & TMean      & $\underline{85.63}$ & $68.76$ & $26.01$ & $\textbf{22.31}$ & $\underline{78.65}$ & $65.94$& $\underline{31.03}$ & $\underline{31.00}$ & $69.23$ & $53.69$ & $39.18$ & $39.67$\\
     \cite{tan2019efficientnet}& Median     & $84.70$ & $\underline{69.52}$ & $21.00$ & $24.05$ & $76.10$ & $57.52$& $38.75$ & $34.27$ & $64.16$ & $37.21$ & $57.25$ & $45.46$  \\   
     & FoolsGold  & $85.23$ & $66.45$ & $36.71$ & $25.39$ & $76.13$ & $56.83$& $56.63$ & $34.81$ & $\underline{70.34}$ & $52.70$ & $\underline{28.70}$ & $\underline{38.25}$ \\
     & FLAME & $82.24$ & $56.09$ & $\underline{11.72}$ & $25.66$ & $77.01$ & $\underline{66.85}$ & $32.57$ & $33.92$ & $69.71$ & $\underline{55.74}$ & $30.43$ & $38.91$ \\
     \rowcolor{lightgray}
     \cellcolor{white}& FLARE (Ours)      & $\textbf{86.20}$ & $\textbf{74.28}$ & $\textbf{6.76}$ & $\underline{23.20}$ & $\textbf{78.95}$ & $\textbf{69.18}$& $\textbf{31.02}$ & $\textbf{30.45}$ & $\textbf{71.58}$ & $\textbf{56.91}$ & $\textbf{28.42}$ & $\textbf{36.73}$ \\
    \bottomrule
   \midrule

    \multirow{8}{*}{Deit-Tiny} & FedAvg-NA & $87.86$ & $82.90$ & $4.18$ & $21.72$   & $81.29$ & $80.95$& $5.66$ & $26.39$   & $74.06$ & $69.86$ & $8.10$ & $33.48$ \\
    \cmidrule(r){2-14}
     & FedAvg     & $83.79$ & $58.02$ & $49.34$ & $25.06$ & $74.33$ & $51.99$& $56.92$ & $37.50$ & $67.86$ & $44.11$ & $49.13$ & $40.83$ \\
     & Krum       & $70.59$ & $45.83$ & $71.41$ & $44.60$ & $55.44$ & $40.72$& $68.12$ & $57.52$ & $52.09$ & $31.69$ & $70.90$ & $57.48$ \\
     & TMean      & $81.84$ & $48.66$ & $69.80$ & $28.19$ & $75.88$ & $57.49$& $44.92$ & $35.03$ & $68.44$ & $46.28$ & $44.23$ & $\underline{40.31}$\\
     \cite{touvron2021training}& Median    & $\textbf{84.43}$ & $\textbf{62.29}$ & $\textbf{34.59}$ & $\textbf{24.23}$ & $74.83$ & $55.09$& $44.43$ & $36.22$ & $68.17$ & $48.44$ & $\textbf{36.76}$ & $40.68$  \\   
     & FoolsGold  & $83.86$ & $61.60$ & $49.41$ & $25.04$ & $\underline{76.56}$ & $58.76$& $\underline{37.20}$ & $34.11$ & $67.76$ & $45.42$ & $39.45$ & $41.12$ \\
     & FLAME & $81.44$ & $59.02$ & $\underline{36.16}$ & $25.02$ & $75.93$ & $\underline{60.39}$ & $38.95$ & $\underline{33.62}$ & $\underline{68.51}$ & $\underline{48.84}$ & $49.48$ & $41.81$ \\
     \rowcolor{lightgray}
     \cellcolor{white}& FLARE (Ours)   & $\underline{84.37}$ & $\underline{62.10}$ & $40.35$ & $\underline{24.41}$ & $\textbf{77.40}$ & $\textbf{61.64}$& $\textbf{30.05}$ & $\textbf{32.42}$ & $\textbf{68.87}$ & $\textbf{53.61}$ & $\underline{38.33}$ & $\textbf{39.66}$ \\
    \bottomrule
  
  \end{tabular}
\begin{tablenotes} 
	\footnotesize
	\item$^\ast$ \textbf{Bold numbers are the best performance.}
	\item$^\dagger$ \underline{Numbers with underline are the second best values.} 
    \item$^{\ddagger}$ NA denotes No Attack. Others without this symbol are all under attack.
\end{tablenotes}
\end{table*}

\begin{figure*}[t]
\centerline{\includegraphics[width=0.98\textwidth]{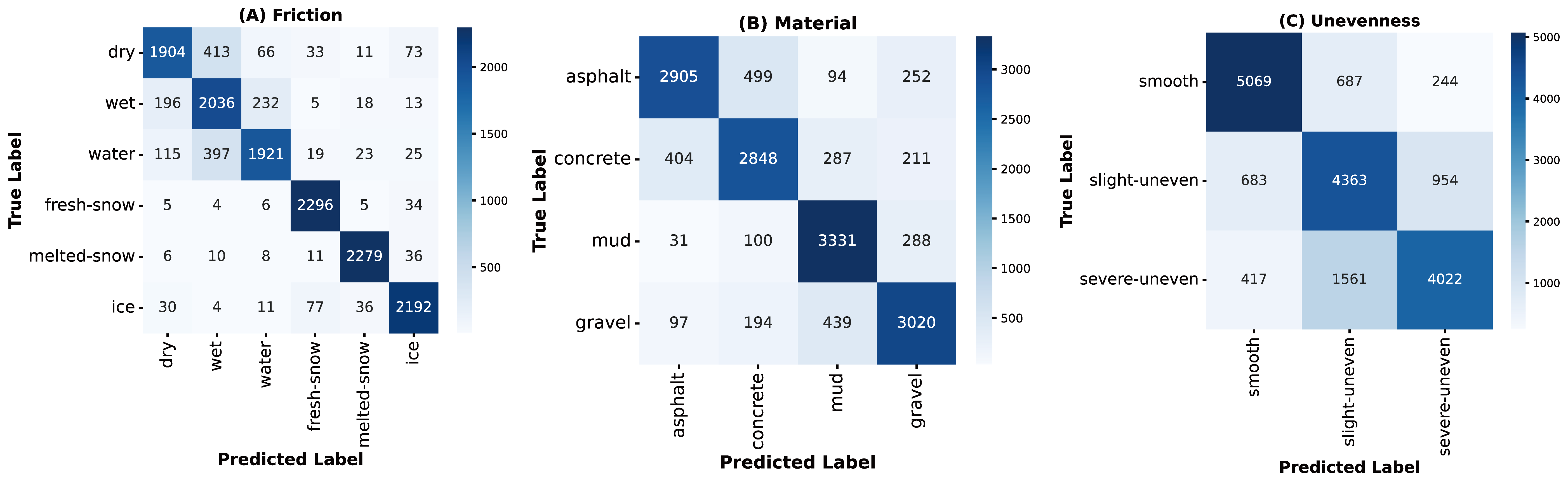}}
\caption{The confusion matrices of FLARE based on EfficientNet-B1 in three RCC tasks: (A) Friction, (B) Material, and (C) Unevenness.}
\label{fig_heatmap}
\end{figure*}

\begin{table*}
  \scriptsize
  \centering
  \caption{Results as a function of poisoned rates based on EfficientNet-B1. All values are ratios in \%.}
  \label{tab_poisoned_rate}
  \renewcommand{\arraystretch}{1.0} 
  \begin{tabular}{cc|cccc|cccc|cccc}
    \toprule
    \multirow{2.5}{*}{Poisoned Rate} & \multirow{2.5}{*}{Method} & \multicolumn{4}{c|}{RCC @ Friction} & \multicolumn{4}{c|}{RCC @ Material}  & \multicolumn{4}{c}{RCC @ Unevenness}\\
    \cmidrule(r){3-14}
    &  & GAcc $\uparrow$ & SRec $\uparrow$& ASR $\downarrow$ & Error $\downarrow$ & GAcc $\uparrow$ & SRec $\uparrow$& ASR $\downarrow$ & Error $\downarrow$ & GAcc $\uparrow$ & SRec $\uparrow$& ASR $\downarrow $ & Error $\downarrow$\\
    \midrule
    - & FedAvg-NA & $87.37$ & $80.18$ & $3.72$ & $19.82$   & $81.36$ & $80.64$& $4.01$ & $26.50$   & $74.45$ & $72.31$ & $9.68$ & $32.97$ \\
    \bottomrule
   \midrule
    \multirow{6}{*}{20\%} & FedAvg & $85.41$ & $67.99$ & $25.11$ & $22.65$ & $79.33$ & $69.41$& $27.20$ & $29.78$ & $71.18$ & $60.46$ & $30.68$ & $37.21$ \\
     & Krum      & $72.70$ & $64.03$ & $\underline{9.01}$  & $40.96$ & $65.82$ & $47.39$& $59.21$ & $47.03$ & $56.31$ & $31.98$ & $57.13$ & $53.19$ \\
     & TMean     & $\underline{86.12}$ & $\underline{72.49}$ & $16.32$ & $\underline{21.64}$ & $79.15$ & $69.42$& $27.82$ & $30.08$ & $71.14$ & $56.12$ & $28.94$ & $37.27$\\
     & Median    & $85.01$ & $69.42$ & $23.49$ & $23.67$ & $78.29$ & $67.24$& $\underline{17.67}$ & $30.95$ & $70.86$ & $\textbf{62.29}$ & $\textbf{22.14}$ & $37.29$  \\   
     & FoolsGold & $85.61$ & $69.99$ & $25.03$ & $22.41$ & $\underline{79.85}$ & $\textbf{73.17}$& $20.82$ & $\underline{28.93}$ & $\textbf{71.78}$ & $58.31$ & $24.15$ & $\underline{36.43}$ \\
     & FLAME & $86.02$ & $71.44$ & $13.13$ & $22.15$ & $78.46$ & $71.12$ & $18.92$ & $29.29$ & $70.81$ & $59.42$ & $25.75$ & $36.82$ \\
     \rowcolor{lightgray}
     \cellcolor{white}& FLARE (Ours)    & $\textbf{86.75}$ & $\textbf{77.00}$ & $\textbf{6.72}$  & $\textbf{20.52}$ & $\textbf{80.21}$ & $\underline{72.16}$& $\textbf{14.39}$ & $\textbf{28.47}$ & $\textbf{72.19}$ & $\underline{61.52}$ & $\underline{23.77}$ & $\textbf{35.88}$ \\
     \bottomrule
   \midrule

    \multirow{6}{*}{40\%} & FedAvg & $83.43$ & $54.80$ & $48.60$ & $25.48$ & $74.41$ & $49.54$& $67.63$ & $37.59$ & $64.53$ & $36.36$ & $63.33$ & $45.36$ \\
     & Krum      & $73.47$ & $51.37$ & $65.37$ & $40.00$ & $63.07$ & $31.29$& $70.96$ & $50.56$ & $48.38$ & $15.09$ & $85.76$ & $62.11$ \\
     & TMean     & $82.77$ & $51.35$ & $44.28$ & $26.42$ & $75.11$ & $49.76$& $50.49$ & $36.61$ & $\underline{68.58}$ & $\textbf{52.24}$ & $\underline{44.58}$ & $4\underline{0.24}$\\
     & Median    & $82.87$ & $\underline{58.84}$ & $47.00$ & $26.66$ & $74.67$ & $50.31$& $51.44$ & $36.50$ & $61.80$ & $29.96$ & $75.64$ & $48.26$  \\   
     & FoolsGold & $\underline{83.61}$ & $57.06$ & $43.18$ & $25.37$ & $77.09$ & $57.46$& $35.76$ & $\underline{33.51}$ & $66.17$ & $40.27$ & $45.49$ & $43.50$ \\
     & FLAME & $83.51$ & $56.39$ & $\underline{34.73}$ & $\underline{25.29}$ & $\underline{77.66}$ & $\underline{62.76}$ & $\underline{32.95}$ & $34.46$ & $65.43$ & $43.43$ & $46.79$ & $42.22$ \\
     \rowcolor{lightgray}
     \cellcolor{white}& FLARE (Ours)    & $\textbf{84.38}$ & $\textbf{62.37}$ & $\textbf{16.80}$ & $\textbf{23.98}$ & $\textbf{79.57}$ & $\textbf{72.83}$& $\textbf{21.12}$ & $\textbf{29.71}$ & $\textbf{68.79}$ & $\underline{45.61}$ & $\textbf{31.62}$ & $\textbf{40.17}$ \\ 
   \midrule  
  \end{tabular}
\end{table*}

Similar to the literature, e.g., \cite{jebreel2024lfighter}, this paper adopts the Dirichlet distribution to create Non-IID training data for each client and the default $\alpha$ is 1.0 in the experiments. By default, there are 12 malicious clients among the 40 preselected (30\% poisoned rate). If included in the FL process, they execute the TLFAs by shifting the labels in their training datasets from \emph{water} to \emph{dry} for Friction, from \emph{gravel} to \emph{asphalt} for Material, and from \emph{severe-uneven} to \emph{smooth} for Unevenness. As for testing datasets, they remain unchanged during the training and are only used for inference.

To evaluate the generality and compatibility of each method, we train three popular DNN models under FL for the RCC tasks: ResNet-18 \cite{he2016deep}, EfficientNet-B1 \cite{tan2019efficientnet}, and Deit-Tiny \cite{touvron2021training}. They are all lightweight versions of DNN thus suitable for vehicles, and the detailed model information is summarized in Table \ref{tab_model}. To evaluate current countermeasures against TLFAs in RCC, we compare the following seven methods in the experiments: FedAvg \cite{mcmahan2017communication} (not really a defense) Krum \cite{blanchard2017machine}, Trimmed Mean \cite{yin2018byzantine}, Median \cite{yin2018byzantine}, FoolsGold \cite{fung2020limitations}, FLAME\cite{nguyen2022flame}, and FLARE. The following four evaluation metrics quantify different performance aspects: Global Accuracy (GAcc), Source Recall (SRec), ASR, and $Error$.

\subsection{TLFA Impact on FL-RCC}

By comparing the results with and without TLFAs, we can analyze the influence of the attack on FL-RCC. Note that NA denotes No Attack, and other rows without this symbol are all under attack. The performance of FedAvg, which is not actually a defense measure, degrades severely compared with FedAvg-NA for the same models and the same RCC tasks. According to Table \ref{tab_results}, we can make the following \textit{assessments}: 1) average reductions caused by the TLFAs to the GAcc among the three models are 3.72\%, 5.05\%, and 5.18\% for Friction, Material, and Unevenness, respectively; 2) the average decrease on SRecs is even more significant, achieving 23.63\%, 23.52\%, and 22.61\% for the three tasks respectively; 3) ASRs of FedAvg increased by 37.93\%, 43.67\%, and 35.76\% compared to FedAvg-NA for the three tasks, respectively; and 4) Errors regarding Friction, Material, and Unevenness grow by 9.01\%, 8.14\% and 6.37\%, respectively. All in all, TLFAs primarily influence ASR and SRec, as the goal of TLFAs is to misguide prediction involving source and target classes.

The influence of the \textit{poisoned rate} can be analyzed based on Table \ref{tab_poisoned_rate} (20\% and 40\% poisoned rate) and EfficientNet-B1's results (30\% poisoned rate) in Table \ref{tab_results}: The higher the poisoned rate, the worse the performance; GAcc, SRec, ASR, and $Error$ all deteriorate gradually when the poisoned rate increases, especially for ASR; whose values degenerate from 3.72\% to 48.60\% for Friction, from 4.01\% to 67.63\% for Material, and from \& 9.68\% to 63.33\% for Unevenness.

From these results, we can further have the following \textit{observations}: 1) in the view of RCC tasks, TLFAs have a more significant impact on Unevenness, which may be because the number of labels in this task is limited while the learned model parameters are more sensitive; 2) from the model standpoint, TLFAs are general and have adverse effects on all evaluated models; and 3) in terms of influence factors, higher poisoned rates one can naturally expect leads to worse performance. In conclusion, current FL-RCC systems are vulnerable to TLFAs, and we should pay more attention to such powerful attacks.   

\subsection{FLARE Defense Effectiveness }

As per Table \ref{tab_results}, the performance of \textit{Krum} is worse than FedAvg in most cases, which may be caused by the fact that Krum simply chooses one local model as the newest global model. Even if Krum can choose the benign one in each round, the advantage of collaboration in FL is gone, thus the performance will degrade. Moreover, once the selected model is the malicious one, the adverse effect is significant. \textit{TMean, Median, FoolsGold, and FLAME} perform better than FedAvg in some cases, especially for SRec and ASR. Considering SRec, the highest values among FedAvg, TMean, Median, FoolsGold, and FLAME are 69.52\% for Friction, 66.85\% for Material, and 55.74\% for Unevenness. However, these values are not satisfactory and unstable. In other words, simply adapting existing countermeasures cannot be effective for the heterogeneous FL-RCC scenarios under TLFAs.

In most cases (88.89\%), \textit{FLARE} can achieve the best or second-best performance compared to the baseline schemes. Specifically, the best ASR values of FLARE are 6.76\% for Friction, 30.05\% for Material, and 28.42\% for Unevenness; the lowest $Error$ values are 23.20\% (Friction), 30.45\% (Material), and 36.73\% (Unevenness). Compared to the best baseline results, on average, 2.23\% and 0.76\% improvements are achieved for ASR and $Error$, respectively. This superiority by FLARE indicates that paying attention to source and target neurons rather than coarsely analyzing the entire model or output layer can indeed improve defense effectiveness.

Based on Table \ref{tab_poisoned_rate}, FLARE is more robust and performs better than other baselines when the \textit{poisoned rate} increases. Specifically, 1) the ASR of FLARE in Friction ranges from 6.72\% to 16.80\%, while the ASR of the best baseline schemes ranges from 9.01\% to 34.73\%; and 2) FLARE can achieve 72.83\% SRec in Material with 40\% poisoned rate, while the values of baseline schemes range from 31.29\% to 62.76\%. These results suggest that the poisoned rate can influence performance severely, while FLARE is more robust when TLFAs are serious. Fig. \ref{fig_heatmap} provides detailed prediction results of FLARE via \textit{confusion matrices}. FLARE can predict most samples correctly, even for those with source classes (water, grave, and severe-uneven). Moreover, most of the misclassified results are concentrated around the true labels, indicating the influence of the poisoning attack is largely mitigated. 

\textit{Discussion:} Although FLARE can achieve overall the best performance compared to baselines for the FL process, two limitations are observed from the results: 1) there is still a huge gap between FLARE and FedAvg-NA; and 2) the number of misclassified samples is not small enough, according to Fig. \ref{fig_heatmap}. These observations indicate that 1) TLFAs are powerful attacks on FL-RCC systems, very difficult to defend against; and 2) filtering out potentially malicious model parameters in aggregation is not enough, and mitigating the deterioration of already poisoned global model could also be beneficial. 

\section{Conclusions}\label{sec_conclusion}

This paper investigates and reveals the vulnerability of current FL-RCC systems to TLFAs. We propose an evaluation metric based on label distance to better reflect transportation safety and a scheme named FLARE based on neuron-wise analysis to better quantify and address such a vulnerability. Experimental results based on various RCC tasks, models, and baselines show the significance of TLFAs and the relative effectiveness of FLARE. Nonetheless, the stability and performance of current countermeasures can be further enhanced. 

\textbf{Ongoing work: } We will investigate TLFA influence across multiple FL-RCC tasks and assess FLARE effectiveness, also measuring latency with embedded hardware. Second, we will further improve FLARE performance, exploring machine unlearning and knowledge distillation for poisoned model correction. Finally, we will consider adaptive colluding adversaries.  

\section*{Acknowledgment}

This work was supported in parts by WASP, VR, and in kind by the KAW Foundation granting access to  Berzelius at the National Supercomputer Centre. 

\bibliographystyle{IEEEtran}
\bibliography{biblio}

\end{document}